\documentclass[pra,twocolumn,showpacs,floatfix]{revtex4}
\usepackage{graphicx}
\begin{document}

\title{Persistent currents in Bose gases confined in annular traps}
\author{S. Bargi$^{1}$, F. Malet$^{1}$, G. M. Kavoulakis$^{2}$, and S.
M. Reimann$^{1}$}
\affiliation{$^{1}$Mathematical Physics, Lund Institute of Technology, 
P.O. Box 118, SE-22100 Lund, Sweden \\
$^{2}$Technological Educational Institute of Crete, P.O. Box 1939,
GR-71004, Heraklion, Greece}
\date{\today}

\begin{abstract}
We examine the problem of stability of persistent currents in 
a mixture of two Bose gases trapped in an annular potential. We 
evaluate the critical coupling for metastability in the transition 
from quasi-one to two-dimensional motion. We also evaluate the critical 
coupling for metastability in a mixture of two species as function of 
the population imbalance. The stability of the currents is shown to be 
sensitive to the deviation from one-dimensional motion.
\end{abstract}

\maketitle

\section{Introduction}

Bose-Einstein condensates of dilute vapors of atoms offer a very 
promising testing ground for questions associated with superfluidity 
for a number of reasons. Firstly, these gases are dilute as 
opposed, for example, to the ``traditional'' superfluid liquid 
helium. Furthermore, the atomic gases can be manipulated in many 
different ways, including the shape of the confining potential, 
the strength and the sign of the effective interatomic interaction, 
the number of different species in multicomponent systems, etc.

While the term ``superfluidity'' covers a whole collection of many 
different phenomena \cite{Leg01}, we focus in the present study on 
the metastability of superflow in annular traps 
\cite{Gupta,Olson,Ryu07,Hen,LK}. Persistent flow has been observed 
recently in a Bose-Einstein condensate of sodium atoms confined in
a toroidal trap \cite{Ryu07}. In this experiment, an initial angular 
momentum of $\hbar$ per particle was transferred to the atoms and 
the rotational flow was observed to persist for up to ten seconds, 
limited only by the trap lifetime and other experimental imperfections. 
Persistent currents with two units of angular momentum were also 
observed in the same experiment.

Theoretical studies have examined the existence of metastable 
states in Bose-Einstein condensed gases trapped in single- 
\cite{Kar07,Smy09,Mas09} and double-ring-like \cite{Mal10} 
confining potentials, as well as other phenomena in double-ring-like
traps \cite{Wolf,Brand}. As shown in Ref.\,\cite{Smy09}, mixtures 
of two components that are confined in a strictly one-dimensional single 
ring support persistent currents, but the interaction strength necessary 
for metastability increases with the admixture of the second component. 
For comparable populations of the two components, the critical value of 
the coupling becomes infinite. Finally, persistent currents with a value 
of the circulation higher than one unit were shown to be stable only 
in single-component systems.

In the present study we investigate the effect of the deviations 
from purely one-dimensional motion on the metastability of the 
currents. We also examine two-component Bose-Einstein condensates, 
and find that it is possible to have metastable superflow for 
sufficiently small admixtures of the second component. One of the 
novel results of our study is that in this latter case the deviation 
from one-dimensional motion gives rise to persistent currents with 
circulation higher than one unit, as opposed to the purely 
one-dimensional case.

The paper is organized as follows. We describe in Sec.~II our 
model, and in Sec.~III the two methods that we have employed to 
solve it, i.e., the mean-field approximation and numerical 
diagonalization of the Hamiltonian. In Sec.~IV we present our 
results. More specifically, in Sec.~IV.A we consider the 
single-component case and study the existence of metastable 
states in the transition from quasi-one-dimensional to 
two-dimensional motion. In Sec.~IV.B we examine the effect 
of the admixture of a second component on the stability of 
the persistent currents, comparing the obtained results with 
the ones of strictly one-dimensional motion. Finally, we 
present our summary and conclusions in Sec.~V.

\section{Model}

\begin{figure}[t]
\centerline{\includegraphics[width=9cm,clip]{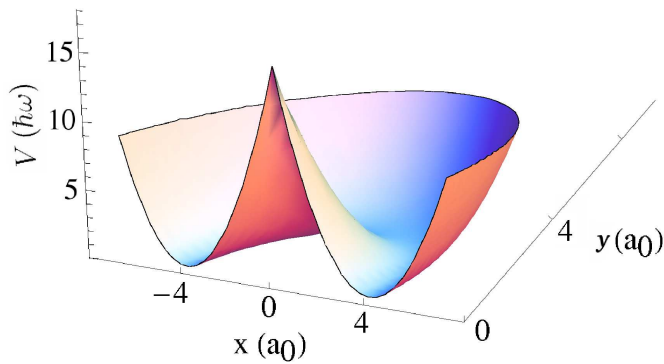}}
\caption{(Color online) The confining potential $V$ of
Eq.\,(\ref{potential}) as a function of the cartesian coordinates 
($x,y$), for $R_{0}/a_0=4$ and $\omega/\omega_{0}=1$.}
\label{fig1} 
\end{figure}

Let us consider two distinguishable species of bosonic atoms, 
labelled as $A$ and $B$, with populations $N_{A}$ and $N_{B}$
respectively; without loss of generality we assume below that 
$N_B \leq N_A$. We also assume two-dimensional motion, which 
in an experiment corresponds to the case of very tight confinement 
in the perpendicular direction. In the plane of motion we model 
the annular trap by a displaced harmonic potential, 
\begin{equation}
  V(\rho)= \frac{1}{2} M \omega^{2} (\rho - R_{0})^{2},
\label{potential}
\end{equation}
which is plotted in Fig.~\ref{fig1}. Here $\rho$ is the usual 
radial variable in cylindrical polar coordinates, $R_{0}$ 
is the radius of the annulus, $\omega$ is the trap frequency,
and $M$ is the atom mass, which we assume for simplicity to 
be equal for both species. We stress that the results presented 
below with the potential of Eq.\,(\ref{potential}) are -- at least 
qualitatively -- similar to the ones that we have obtained using 
a harmonic-plus-Gaussian trapping potential.

In order to investigate the effects due to the transition from
quasi-one-dimensional to two-dimensional motion, we vary 
the confinement strength $\omega$ in the radial direction. For 
large $\omega$, the confinement is strong enough to freeze out 
the motion in this direction, which makes the system 
quasi-one-dimensional. As $\omega$ decreases, the confinement 
becomes weaker and, as a result, the motion becomes two-dimensional. 
In all the calculations presented below we fix the radius of the 
annulus to $R_{0}/a_0= 4$, where $a_{0}=\sqrt{\hbar/(M\omega_{0})}$ 
is the ususal oscillator length corresponding to some fixed 
frequency $\omega_0$.

The interatomic interactions are modelled via the usual effective 
contact potential, which is assumed to be repulsive. Therefore, 
the Hamiltonian $H$ of the system is $H = H_{\rm sp} + H_{\rm int}$,
where $H_{\rm sp}$ is the single-particle part 
\begin{eqnarray}
  H_{\mathrm{sp}} & = & 
 \sum_{i=1}^{N}\left(-\frac{\hbar^{2}}{2M}\nabla_{i}^{2}
 + V(\mathbf{r}_{i})\right),
\label{Hsp}
\end{eqnarray}
and $H_{\rm int}$ is the interaction part, given by 
\begin{eqnarray}
  H_{\mathrm{int}} = 
\frac{u_{AA}}{2}\sum_{i\neq j=1}^{N_{A}}
\delta({\bf r}_{i}-{\bf r}_{j})
+\frac{u_{BB}}{2}\sum_{i\neq j=1}^{N_{B}}
\delta({\bf r}_{i}-{\bf r}_{j})
\nonumber \\
+u_{AB}\sum_{i=1,j=1}^{N_{A},N_{B}}\delta({\bf r}_{i}-{\bf r}_{j}).
\label{Hint}
\end{eqnarray}
Here the parameters $u_{kl}$ are proportional to the s-wave scattering 
lengths for zero-energy elastic atom-atom collisions. For simplicity, 
we take $u_{AA}=u_{BB}=u_{AB}\equiv u$, which is also experimentally 
relevant in several cases; furthermore, since the interaction is assumed 
repulsive, $u>0$.

\section{Method}

We attack this problem using both the mean-field Gross-Pitaevskii
approximation, as well as diagonalization of the many-body Hamiltonian.

Within the diagonalization approach, we fix the particle numbers $N_{A}$ 
and $N_{B}$ and the total angular momentum $L \hbar$, and use the Lanczos 
method \cite{Lan50} to obtain the eigenenergies and the corresponding 
eigenvectors. This can be done for a range of the values of $L$, 
yielding the dispersion relation $E(L)$, i.e., the smallest eigenvalue 
for some given $L$, allowing us to identify the possible local minima 
associated with the metastable states that give rise to persistent currents 
in the system. 

The basis states that we choose in this approach are the eigenfunctions 
of the single-particle problem, which are of the form 
\begin{equation}
  \psi_{n,m}(\rho,\theta)=\frac{e^{i m \theta}}{\sqrt{2\pi}} R_{n,m}(\rho).
\end{equation}
The quantum number $n$ corresponds to the radial excitations, $m$ 
is the quantum number associated with the angular momentum, which 
results from the rotational symmetry of our problem, and $\theta$ 
is the usual angular variable in cylindrical polar coordinates. The 
radial wavefunctions $R_{n,m}(\rho)$ are evaluated numerically. If 
$E_{n,m}$ are the corresponding eigenvalues, we work under the 
assumption that the typical value of the interaction energy 
$V_{\rm int}$ is weak enough, so that $V_{\rm int} \ll E_{1,0}$. 
Thus, we set $n = 0$ and the only quantum number that remains is 
$m$, whose highest value is chosen so that $V_{\rm int} \lesssim 
E_{0,m} \ll E_{1,0}$. Having evaluated the single-particle states, 
these are then combined in all possible ways to form a basis of the 
Fock states for the many-body problem for some fixed $N_A$, $N_B$, 
and $L$. 

Turning to the mean-field approximation, within this approach the 
system is described via the two order parameters of the two components, 
$\phi_{A}$ and $\phi_{B}$. From Eqs.~(\ref{Hsp}) and (\ref{Hint}) one 
obtains the two coupled Gross-Pitaevskii-like equations for these two 
order parameters 
\begin{eqnarray}
 -\frac{\hbar^{2}\nabla^{2}}{2M}{\phi}_{A}+
V(r){\phi}_{A}+g_{AA}|{\phi}_{A}|^{2}{\phi}_{A}+
g_{AB}|{\phi}_{B}|^{2}{\phi}_{A}\nonumber \\
={\mu}_{A}{\phi}_{A},
\nonumber \\
-\frac{\hbar^{2}\nabla^{2}}{2M}{\phi}_{B}+
V(r){\phi}_{B}+g_{BB}|{\phi}_{B}|^{2}{\phi}_{B}+
g_{AB}|{\phi}_{A}|^{2}{\phi}_{B}
\nonumber \\
={\mu}_{B}{\phi}_{B}.
\label{gpe}
\end{eqnarray}
In the above equations the two order parameters are normalized as 
$\int|{\phi}_{A}|^{2} d^2{\bf r}=1$, and $\int|{\phi}_{B}|^{2}d^2{\bf r}=
N_B/N_A$. The parameters $g_{ij}$ are defined as $g_{AA}=N_A u_{AA}$, 
$g_{BB}=N_A u_{BB}$, and $g_{AB}=N_A u_{AB}$. Finally, $\mu_{A}$ and 
$\mu_B$ are the chemical potentials of the two components. 

To solve Eqs.~(\ref{gpe}) we make use of a fourth-order split-step
Fourier method within an imaginary-time propagation approach 
\cite{Chi05}. Starting with an initial state with some given angular 
momentum, the imaginary-time propagation drives the system along 
the dispersion relation $E(L)$ until it finds a local minimum. If 
$E(L)$ has such local minima, the final state may be a metastable 
one. Otherwise, the initial state decays to the non-rotating ground 
state of the system. One should note that the dissipation of the 
angular momentum $L$ is inherent to the diffusive character of this 
method, and it does not contradict the fact that the Hamiltonian 
commutes with the operator of the angular momentum, since the 
imaginary-time propagation operator is not unitary, and therefore 
does not satisfy Heisenberg's equation of motion \cite{Bra89}.

The two methods mentioned above complement each other: the 
diagonalization gives the full dispersion relation $E(L)$ for 
any fixed angular momentum $L \hbar$, but due to numerical 
limitations in diagonalizing matrices of large size, one has to 
restrict the number of particles, as well as the values of the 
interaction strength. The mean-field Gross-Pitaevskii approach, 
on the other hand, does not have these limitations, but does not 
allow us to evaluate the whole dispersion relation $E(L)$ -- at 
least in a straightforward way, but rather its local/absolute minima. 
Finally, due to its mean-field character, possible correlations 
between the atoms in certain limiting cases are not captured 
within this approximation, although they are not expected to be 
important in the problem that we consider here. 

\section{Results}

\subsection{Deviation from one-dimensional motion and metastability}

\begin{figure}
\centerline{\includegraphics[width=9cm,clip]{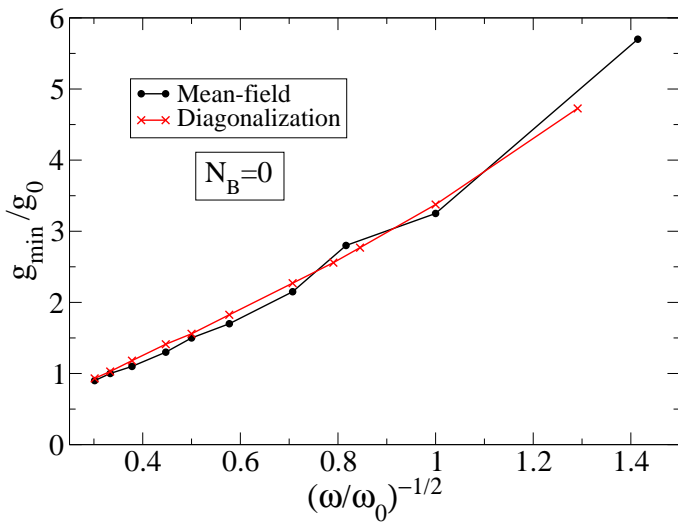}}
\caption{(Color online) Minimum interaction strength 
$g_{\mathrm{min}}/g_0$ for the existence of persistent currents, 
as a function of $(\omega/\omega_{0})^{-1/2}$ in an annular 
trap with $R_{0}/a_0 = 4$. The (black) circles show the result 
from the mean-field approximation, and the (red) crosses from 
the diagonalization of the Hamiltonian.}
\label{fig2}
\end{figure}

We start with the case of a single component, i.e., when the total 
number of atoms $N = N_A + N_B$ is equal to $N_{A}$. We evaluate 
the minimum interaction strength $g_{\mathrm{min}}$ necessary for 
the system to support persistent currents as a function of the 
width of the annulus, or equivalently as a function of the confinement 
frequency $\omega$. As we argued earlier, for large values of $\omega$, 
the motion becomes quasi-one-dimensional, while as $\omega$ decreases, 
the motion becomes two-dimensional. We solve Eqs.~(\ref{gpe}) and 
also diagonalize the Hamiltonian numerically for several values of 
$\omega$. The results of both calculations are shown in Fig.~\ref{fig2}, 
represented by (black) circles (mean-field) and (red) crosses 
(diagonalization). As seen from this plot, the results obtained from 
the two approaches are in good agreement with each other.

At this point it is instructive to see how these data compare with 
the analytical result of a purely one-dimensional model. To reduce 
the two-dimensional problem into an effectively one-dimensional one, at 
least in the limit where the width of the annulus (set by the oscillator 
length) is much smaller than its radius $R_0$, one may start from the 
initial expression for the energy and integrate over the radial degrees 
of freedom \cite{JKP}. To simplify this calculation, we assume that the 
density in the transverse direction is nonzero only for $R_0 - 
a_{\rm osc}/2 \le \rho \le R_0 + a_{\rm osc}/2$, where $R_0$ is the 
radius of the annulus and $a_{\rm osc}$ is the oscillator length 
corresponding to $\omega$. In this limit, the order parameter may 
be written in a product form,
\begin{equation}
 \phi_A(\rho, \theta) = R_A(\rho) \, \Theta_A(\theta).
\label{prod}
\end{equation}
Assuming the above factorization, one may start from the 
two-dimensional problem and derive the following one-dimensional
nonlinear equation 
\begin{equation}
 - \frac {\hbar^2} {2 M R_0^2} \frac {\partial^2 \Theta_A(\theta)} 
{\partial \theta^2} + \frac g {a_{\rm osc} R_0} |\Theta_A(\theta)|^2 
\Theta_A(\theta) = \mu_A \Theta_A(\theta),
\label{prodd}
\end{equation}
where the coefficient of the nonlinear term involves the integral
$\int |R_A(\rho)|^4 \, \rho d \rho$. According to previous studies 
\cite{Han,GK}, Eq.\,(\ref{prodd}) implies that the minimum value 
of the coupling $g$ for metastability of currents with one unit of
circulation, i.e., with $2 \pi \hbar/M$, is (in the limit $R_0 \gg 
a_{\rm osc}$)
\begin{equation}
  g_{\rm min} \approx \frac {3 \pi} 2 \frac {\hbar^2} M 
\frac {a_{\rm osc}} {R_0} = \frac {3 \pi} 2 \frac {\hbar^{5/2}} {M^{3/2}} 
\frac {1} {R_0 \sqrt{\omega}}\,,
\label{gmin}
\end{equation}
i.e., it scales as $\omega^{-1/2}$. Clearly the numerical prefactor 
in the above expression depends on the form of the order parameter in 
the transverse direction $R(\rho)$. If we define $g_0 \equiv \hbar^{5/2}
/(M^{3/2} \omega_0^{1/2} R_0)$, this expression may be written in the 
more convenient form
\begin{equation}
  \frac {g_{\rm min}} {g_0} = \frac {3 \pi} 2 
\sqrt{ \frac {\omega_0} {\omega}}\,.
\label{gminn}
\end{equation}
Within this simplified model, the obtained power-law dependence 
of $g_{\rm min}$ scaling as $\omega^{-1/2}$ is in agreement with 
the numerical results plotted in Fig.\,2. This figure shows that 
$g_{\rm min}$ indeed increases linearly as a function of 
$(\omega/\omega_{0})^{-1/2}$ for strong confinement in the 
transverse direction, when the motion is quasi-one-dimensional. 
From the numerical data we also see that for $\omega/\omega_{0} 
\lesssim 1$, $g_{\rm min}$ grows faster, as the extra degrees of 
freedom associated with the motion in the transverse direction 
start to play a role. We also notice that for small values of 
$\omega$, when the confinement becomes weak, the maximum value 
of the confining potential at the center of the trap decreases. 
For example, for $\omega/\omega_{0} \sim 0.25$, the density of 
the system is already substantial at the center of the trap,
and the deviations from quasi-one-dimensional motion are
very pronounced.

\subsection{Population imbalance and metastability}

We now turn to the question of metastability in a two-component
system. Again, we study the minimum interaction strength for the
existence of metastable flow when the atoms are confined in a
(two-dimensional) annular trap. To do this, we take $R_0/a_0= 4$ 
as before, and consider a fixed and relatively weak confinement 
$\omega/\omega_{0}=1$, varying $x_A \equiv N_A/(N_{A}+N_{B})$. 
This can be done easily within the Gross-Pitaevskii scheme, where
one can choose any value for the relative population $N_B/N_A$, 
but not within the diagonalization approach, where one has to 
specify the (integer) particle numbers $N_A$ and $N_B$, and 
therefore the values of $x_A$ are more restricted. In addition, 
in order to achieve small steps in $x_A$, the number of particles 
has to be fairly large, which results in Hamiltonian matrices 
of large size, making it hard to diagonalize numerically.

As before, one can obtain a simple analytic result for $g_{\rm min}$ 
from the result of a strictly one-dimensional trapping potential 
\cite{Smy09}. Following the same arguments as in the previous 
subsection, then Eq.\,(\ref{prodd}) implies that, with the
assumption of a step function for the density of the gas
in the transverse direction,
\begin{equation}
  \frac {g_{\rm min}} {g_0} = \frac{3 \pi} {2 (4 x_A - 3)} 
 \sqrt{ \frac {\omega_0} {\omega}},
\label{eq:g_min_2comp}
\end{equation}
which clearly reduces to Eq.\,(\ref{gminn}) when $x_A = 1$. 
According to this simplified model, $g_{\rm min}$ diverges 
at $x_A = 3/4$ and thus metastability cannot exist for $x_A 
\le 3/4$. 

\begin{figure}
\centerline{\includegraphics[width=9.5cm,clip]{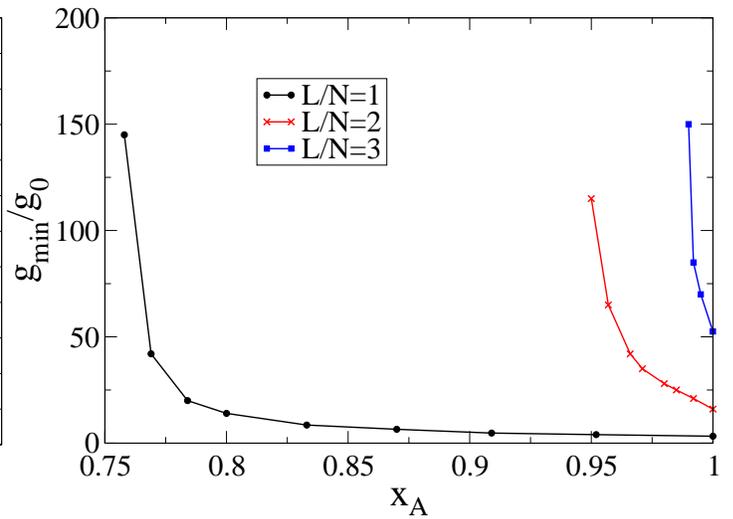}}
\caption{(Color online) The minimum interaction strength 
$g_{\mathrm{min}}/g_0$ for stability of persistent currents 
in an annular trap with $R_{0}/a_0 = 4$ and $\omega/\omega_{0}=1$, 
as a function of the relative population $x_A$, obtained within 
the mean-field approximation. The (black) circles correspond to
$L/N = 1$, the (red) crosses to $L/N = 2$, and the (blue)
squares to $L/N = 3$.}
\label{fig3} 
\end{figure}

In Fig.~\ref{fig3} we show the numerical results that we obtain 
for the minimum interaction strength necessary for the existence 
of metastable states as a function of $x_A$, within the mean-field 
approximation. The (black) circles correspond to a value of the
angular momentum per particle $L/N$ equal to unity. Our calculations 
indicate a divergence of $g_{\rm min}$ for sufficiently small values 
of $x_A$, pretty much like the one-dimensional case, where this 
divergence occurs at $x_A=3/4$. On the other hand, metastability 
takes place at a stronger interaction strength than in the purely 
one-dimensional case. Therefore, the deviation from this limit works 
against the stability of the currents. This result is consistent 
with the general statement that metastability of the superflow is 
not possible in trapping potentials which decrease monotonically 
with the distance from the center of the trap \cite{met}.

What has been mentioned so far has to do with values of the 
circulation equal to one unit, i.e., equal to $2 \pi \hbar/M$. 
According to Ref.\,\cite{Smy09}, in the strictly one-dimensional 
problem persistent currents with circulation higher than one unit 
are stable only in the single-component case, i.e., for $x_A=1$ 
(or equivalently $N_B = 0$). Interestingly, unlike the one-dimensional 
case, our mean-field calculations show that persistent currents 
with a value of the circulation which is higher than one unit are 
stable for finite admixtures of a second component. This result 
is shown in Fig.\,3 in the two higher curves, represented as (red) 
crosses, and (blue) squares, corresponding to $L/N=2$ and $L/N=3$, 
respectively.

The two limiting cases of strong and weak confinement are worth 
commenting on: as one approaches the one-dimensional limit, i.e., 
when $\omega$ increases, the two curves corresponding to higher 
values of the circulation approach infinity more rapidly, eventually 
becoming vertical, as in the purely one-dimensional case. Thus, 
metastability with $L/N=2$ or $L/N=3$ is not possible in this limit, 
in agreement with the results of Ref.\,\cite{Smy09}. Metastability 
is not possible, when the confinement becomes very weak, either, 
i.e., when the width of the annulus gets large, and the gas has 
a significantly nonzero density at the center of the trap. In 
other words, the two curves have an infinite slope in the two 
limiting cases (of small and large values of $\omega$), but have 
a finite slope for intermediate values, indicating the stability 
of persistent currents in this regime.

While the behavior of the system looks similar in both limits,
the underlying physical origin of the non-monotonic behavior
of the slope of these curves is different in the two extremes.
In the limit of small $\omega$, this is due to the fact that
metastability is absent in an almost homogeneous gas \cite{met}. 
In the opposite limit of large $\omega$, it is the one-dimensionality
of the problem that causes this effect \cite{Smy09}.

\begin{figure}
\includegraphics[width=9cm,clip]{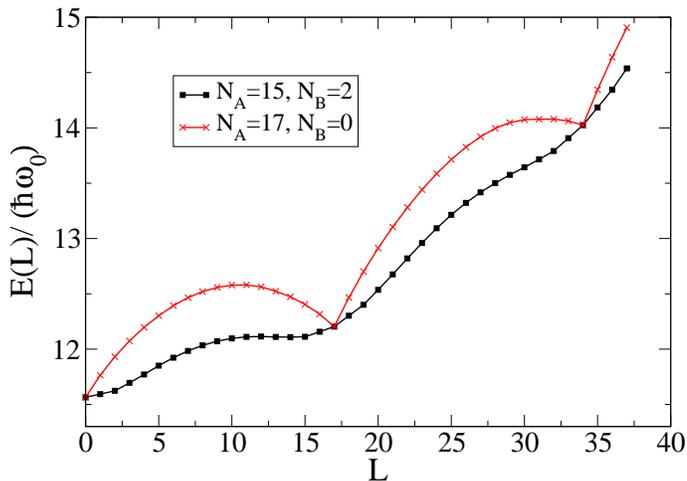}
\caption{(Color online) The dispersion relation $E(L)$ obtained
from the diagonalization of the Hamiltonian, in a (two-dimensional)
annular trap, with $N = 17$ atoms, $\omega/\omega_{0}=1$, $R_{0}
/a_0=4$, and $g/g_0=30$. The (red) crosses show the single-component 
case, with $N_A = 17$ and $N_B = 0$ ($x_A = 1$), while the (black) 
squares correspond to a two-component system, with $N_A = 15$ and 
$N_B =2$ ($x_A \approx 0.88$). While in the first case there are 
two local minima in $E(L)$ at $L/N=1$ and $L/N=2$, in the second 
case there is only one minimum, around $L/N=1$.}
\label{fig4}
\end{figure}

The results that we have obtained from the numerical diagonalization 
of the Hamiltonian are shown in Fig.~\ref{fig4} and are consistent 
with those of the mean-field approximation. Figure \ref{fig4} shows 
the dispersion relation, i.e., the lowest eigenvalue of the Hamiltonian 
for each value of $L$, with $x_A=1$ (red crosses) and $x_A \approx 0.88$ 
(black squares). More specifically, we have considered $N=17$ atoms, 
with $N_A = 17$ and $N_B = 0$ in the one case, and $N_A = 15$ and $N_B = 2$ 
atoms in the other one. In this calculation the single-particle basis 
includes all the states with $\left|m\right| \le 5$, and we consider 
the same parameters as before, with $R_0/a_0=4$, $\omega/\omega_{0}=1$, 
and $g/g_0=30$. 

The dispersion relation $E(L)$ in the case of a single component 
has two local minima, at $L/N = 1$ and $L/N = 2$, showing that 
persistent currents are stable in both cases. This is consistent 
with the result obtained from the mean-field calculations shown in 
Fig.\,3. On the other hand, in the case of two components, $E(L)$ has 
only one local minimum around $L/N=1$. Again, this is consistent with 
the result of Fig.\,3, since for the corresponding values of $x_A$ and 
$g_{\rm min}/g_0$, only the state with $L/N = 1$ is metastable.

\section{Summary and conclusions}

According to the present study, Bose-Einstein condensed atoms 
which move in annular traps offer a very interesting system for 
the study of persistent currents. 

In principle one could investigate a surface in a three-dimensional 
phase diagram showing the minimum interaction strength for stable 
currents as function of both the confinement strength $\omega$,
which determines the deviation from purely-one-dimensional motion, 
and the population imbalance $x_A$. Given the size and the numerical 
effort of such a calculation, we have instead restricted ourselves to 
certain parts of this diagram, and in particular we have investigated 
two different questions.

Firstly, we considered the case of one component, and showed that as 
the width of the annulus becomes larger, and thus there are deviations 
from the strictly one-dimensional motion, the critical coupling 
$g_{\rm min}$ that is necessary to achieve metastability increases. 
This result is consistent with the fact that a necessary condition for 
metastability is that the confining potential -- and as a result the 
particle density -- does not decrease monotonically from the center of 
the trap; in the picture of vortex dynamics, there is a force acting on 
the vortex that is associated with the rotational motion of the superfluid, 
which is in the opposite direction of the gradient of the single-particle 
density distribution. 

Secondly, we investigated the case of a mixture of two components, 
in a fixed annular potential. We found that $g_{\rm min}$ which corresponds 
to persistent currents with one unit of circulation increases with the 
addition of the second component. Also, we found a similar behavior as 
in the one-dimensional problem, where $g_{\rm min}$ diverges at $x_A = 3/4$. 
Thus, the main effect of the finite width of the annulus is a relative 
increase in the critical coupling.
 
On the other hand, a novel result that is absent in the one-dimensional 
case is the stability of persistent currents with circulation for the 
larger component which is higher than one unit. When the motion is 
one-dimensional, $g_{\rm min}$ is infinite, and there are no stable 
currents. As the width of the annulus increases, these currents
become stable for interaction strengths which are sufficiently 
strong, but finite. When the annulus becomes even wider, $g_{\rm min}$
becomes infinite again. Therefore, the effect of the deviation 
from the one-dimensional motion is rather dramatic in this case.

From the above results it is clear that the phase diagram $g_{\rm min}
= g_{\rm min}(\omega, x_A)$ has an interesting structure and that 
the physics of persistent currents of Bose-Einstein condensed gases 
trapped in annular potentials is very rich. With the recent progress 
in building such trapping potentials in the laboratory, it will be of 
much interest to investigate all these effects also experimentally.

\section{Acknowledgements}

This work was financed by the Swedish Research Council. The collaboration
is part of the NordForsk Nordic network ``Coherent Quantum Gases - From 
Cold Atoms to Condensed Matter''. GMK acknowledges useful discussions 
with M. Magiropoulos and J. Smyrnakis.

\end{document}